\documentclass[final]{IEEEtran}
\usepackage{amsthm,amssymb,graphicx,multirow,amsmath,color,amsfonts}
\usepackage[update,prepend]{epstopdf}
\usepackage{algorithm}
\usepackage[noend]{algpseudocode}
\usepackage[noadjust]{cite}
\usepackage[latin1]{inputenc}
\usepackage{tikz}
\usepackage{pdfpages}
\usepackage{subfigure,array}
\usepackage{tabulary}
\usepackage{multirow}
\usepackage{float}
\usepackage{filecontents}
\usepackage{subfigure}
\usepackage[font=footnotesize]{caption}
\usepackage{comment}

\usepackage{mathtools}
\let\subparagraph\relax
\usepackage[compact]{titlesec}
\titlespacing{\section}{0pt}{*1}{*1}
\titlespacing{\subsection}{0pt}{*1}{*1}

\let\emptyset\varnothing
\newcommand\norm[1]{\left\lVert#1\right\rVert}

\def\nb0{{\mathbf{0}}}
\def\nb1{{\mathbf{1}}}

\newtheorem{lemma}{Lemma}

\newtheorem{ndef}{Definition}

\newtheorem{theorem}{Theorem}

\newtheorem{remark}{Remark}

\def\argmin{\operatorname{arg~min}}
\def\argmax{\operatorname{arg~max}}
\def\E{\mathbb{E}}

\def\R{\mathbb{R}}

\def\orthvect{{\cal U}(M_t,1)}
\def\grvect{{\cal G}(M_t,1)}
\def\compvect{{\mathbb C}^{M_t \times 1}}
\def\C{{\mathbb C}}
\def\compchan{{\mathbb C}^{M_r \times M_t}}

\allowdisplaybreaks 
\title{Learning on a Grassmann Manifold: \\
CSI Quantization for Massive MIMO Systems}

\author{
Keerthana Bhogi, Chiranjib Saha, and Harpreet S. Dhillon
\thanks{K. Bhogi, C. Saha, and H. S. Dhillon are with Wireless@VT, Department of ECE, Virginia Tech, Blacksburg, VA, USA. Email: \{kbhogi, csaha, hdhillon\}@vt.edu. The support of the US National Science Foundation (Grants ECCS-1731711 and CNS-1923807)  is gratefully acknowledged. 
} }

\begin{document}
\maketitle
\begin{abstract}
This paper focuses on the design of beamforming codebooks that maximize the average normalized beamforming gain for any underlying channel distribution. While the existing techniques use statistical channel models, we utilize a model-free data-driven approach with foundations in machine learning to generate beamforming codebooks that adapt to the surrounding propagation conditions. The key technical contribution lies in reducing the codebook design problem to an unsupervised clustering problem on a Grassmann manifold where the cluster centroids form the finite-sized beamforming codebook for the channel state information (CSI), which can be efficiently solved using $K$-means clustering. This approach is extended to develop a remarkably efficient procedure for designing product codebooks for full-dimension (FD) multiple-input multiple-output (MIMO) systems with uniform planar array (UPA) antennas. Simulation results demonstrate the capability of the proposed design criterion in learning the codebooks, reducing the codebook size and producing noticeably higher beamforming gains compared to the existing state-of-the-art CSI quantization techniques.
\end{abstract}
\begin{IEEEkeywords}
Massive MIMO, full-dimension (FD) MIMO, FDD, beamforming, codebook, machine learning, Grassmann manifold, $K$-means clustering.
\end{IEEEkeywords}

\section{Introduction}
Transmit beamforming with receive combining is one of the simplest approaches to achieve full diversity in a MIMO system. It just requires CSI at the transmitter in the form of the transmit beamforming vector. The frequency division duplex (FDD) large-scale MIMO systems cannot utilize channel reciprocity to acquire CSI at the transmitter using uplink transmission. This necessitates channel estimation using downlink pilots and the subsequent feedback of the channel estimates (in this case the beamforming vector) to the transmitter over a dedicated feedback channel with limited capacity. This results in a significant overhead when the number of antennas is large. One way to overcome this problem is to construct a set of beamforming vectors constituting a {\em codebook}, which is known to both the transmitter and the receiver. The problem then reduces to determining the {\em best} beamforming vector at the receiver and conveying its {\em index} to the transmitter over the feedback channel~\cite{narula1998efficient}. A key step in this procedure is to construct the codebook, which is a classical problem in MIMO communications~\cite{love2008overview}. A common feature of all classical works in this direction is the assumption of a statistical channel model (such as Rayleigh) for which the {\em optimal} codebooks are constructed to optimize system performance.

With the recent interest in data-driven approaches to wireless system design, it is quite natural to wonder whether machine learning has any role to play in this classical problem. Since the fundamental difficulty in this problem is the dimensionality of the channel, the natural tendency is to think in terms of obtaining a low dimensional representation of the channel using deep learning techniques, such as autoencoders~\cite{deepCSI2018}, \cite{deepCSI2019}, which can be used for codebook construction. An autoencoder operates on the hypothesis that the data possesses a representation on a lower dimensional manifold of the feature space, {\em albeit} unknown, and tries to learn the embedded manifold by training over the dataset. In contrast, for MIMO beamforming, the underlying manifold is known to be a Grassmann manifold. This removes the requirement of ``learning'' the manifold from the dataset which often times can be extremely complicated. Once the manifold is known, we can leverage the ``shallow'' learning techniques like the clustering algorithms on the manifold to find the optimal codebook for beamforming. 
 
{\em Prior Art.} As is the case with any communication theory problem, almost all existing works on limited feedback assume {\em some} analytical model for the channel to enable tractable analyses, e.g., i.i.d Rayleigh fading~\cite{mukkavilli2003beamforming}, spatial correlation~\cite{xia2006design}, temporal correlation~\cite{huang2009limited} or both~\cite{lee2015antenna}. Specifically, the problem of quantized maximum ratio transmission (MRT) beamforming can be interpreted as a Grassmannian line packing problem for both uncorrelated~\cite{love2003grassmannian} and spatially correlated~\cite{love2006limited} Rayleigh fading channels and has been extensively studied. The idea of connecting Grassmann manifolds to wireless communications is not new and has been used in other aspects of MIMO systems, such as non-coherent communication~\cite{zheng2002communication} and limited feedback unitary precoding~\cite{love2005limited}. Coming to the context of the limited feedback FDD MIMO, the codebook based on Grassmannian line packing is strictly dependent on the assumption of Rayleigh fading and hence cannot be extended to more realistic scenarios. On the other hand, the discrete Fourier transform (DFT) based beamforming exploits the second order statistics of the channel (such as the direction of departure of the dominant path) and offers a simple yet robust solution to the codebook construction. Owing to the direct connection with the spatial parameters of the channel, the DFT codebook can be extended to the Kronecker product (KP) codebook for 3D beamforming in FD MIMO scenarios. A major drawback of DFT codebook is that it scans all possible directions even though many of them may not be used and thus the available feedback bits are not used efficiently. Finally, since we will be proposing a clustering based solution, it is useful to note that clustering has already found applications in many related problems, such as MIMO detection~\cite{huang2018machine}, automatic modulation recognition~\cite{du2018automatic}, and radio resource allocation in a heterogeneous network~\cite{abdelnasser2014clustering}.

{\em Contributions.} The key technical contribution of this paper lies in the novel formulation of transmit beamforming codebook design for any arbitrary channel distribution as the Grassmannian $K$-means clustering problem. First, we develop the algorithm for $K$-means clustering on the Grassmann manifold that finds the $K$ centroids of the clusters. Leveraging the fact that optimal MRT beamforming vectors lie on a Grassmann manifold, we then develop the design criterion for optimal beamforming codebooks. We then formally establish the connection between the Grassmannian $K$-means algorithm and the codebook design problem and show that the optimal codebook is nothing but the set of centroids given by the $K$-means algorithm. This approach is further extended to develop product codebooks for FD-MIMO systems employing UPA antennas. In particular, we show that under the ${\rm rank}$-$1$ approximation of the channel, the optimal codebook can be decomposed as the Cartesian product of two Grassmannian codebooks of smaller dimensions. We discuss the optimality and performance of the codebooks using both the proposed techniques in terms of average normalized beamforming gain.










{\bf Notation.} We use boldface small case (upper case) letters, e.g. ${\bf a} (\bf A)$, to designate column vectors (matrices) with entries in $\C$. We use $\C^{M \times N}$ to represent $M \times N$ dimensional complex space, ${\cal U}(M,N)$ to represent the set of all $M \times N$ orthonormal matrices, ${\cal U}_M$ to represent the set of all $M \times M$ unitary matrices. Further, $a^* ({\bf a}^*)$ denotes complex conjugate of $a\in \C$ $({\bf a} \in \C^{M \times 1}$), ${\bf A}^{T}$ denotes transpose, ${\bf A}^H$ denotes hermitian, ${\rm svd}({\bf A})$ denotes the singular value decomposition, ${\rm vec}({\bf A})$ denotes the vectorization of a matrix ${\bf A}$. Also, ${\E}_{\bf a}[\cdot]$ represents the expectation over the distribution of ${\bf a}$, $|\cdot|$ denotes the absolute value, $\norm{\cdot}_2$ denotes the matrix two-norm and $j = \sqrt{-1}$. 

\section{System Overview}
We consider a narrow-band point-to-point $M_t\times M_r$ MIMO communication scenario, where the  transmitter and receiver are equipped with $M_t$ and $M_r$ antennas, respectively. In this paper, we focus on the transmit beamforming operation, where the transmitter sends one data stream over a flat fading channel. The discrete-time baseband input-output relation for this system can be expressed as
\begin{align}
    {\bf y} &={\bf {H}}{\bf f}s + {\bf n} , \label{eq::system::model}
\end{align}
where ${\bf y} \in \mathbb{C}^{M_r\times 1}$ is the received baseband signal, ${\bf H} \in \compchan$ is the block fading MIMO channel, $s\in \C$ is the transmitted symbol, ${\bf n}\in{\C}^{M_r\times 1}$ is the additive noise at the receiver, and ${\bf f}\in {\C}^{M_t\times 1}$ is the beamforming vector. The symbol energy is given by $\E[|s|^2] = {\cal E}_t$ and the total transmitted energy is $\E[\norm{{\bf f}s}_2^2] = {\cal E}_t\norm{{\bf f}}_2^2$. The additive noise ${\bf n}$ is Gaussian, i.e., entries in ${\bf n}$ are i.i.d according to $\mathcal{CN}(0,N_o)$. It is assumed that perfect channel knowledge is always available at the receiver. With the combining vector ${\bf z}\in \C^{M_r\times 1}$, the estimated transmitted symbol is obtained as $\hat{s} = {\bf z}^H{\bf y}$. The receive SNR is 
\begin{align}
\gamma_r &=  \frac{{\cal E}_t|{\bf z}^H{{\bf H}{\bf f}}|^2}{|{\bf z}^H{\bf n}{\bf n}^H{\bf z}|}= \gamma_t \frac{| {\bf z}^H{{\bf H}{\bf f}}|^2}{\|{\bf z}\|_2^2\|{\bf f}\|_2^2},\notag
\end{align}
where $\gamma_t={\cal E}_t\|{\bf f}\|^2/N_o$ is the transmit SNR. Without loss of generality, it is assumed that $\norm{{\bf z}}^2_2 = 1$. Under this assumption
\begin{align}
\gamma_r &= {\gamma_t}\frac{{|{\bf z}^H{{\bf H}{\bf f}}|^2}}{\|{\bf f}\|^2} =  {\gamma_t}{\Gamma}({\bf f},{\bf z}), \label{eq::Rx::SNR}
\end{align}
where ${\Gamma}({\bf f},{\bf z})$ 
is the effective channel gain or the beamforming gain. The MIMO beamfoming problem is to choose ${\bf f}$ and ${\bf z}$ such that ${\Gamma}({\bf f},{\bf z})$ is maximized, which would in turn maximize the SNR and consequently minimize the average probability of error and maximize the capacity \cite{andersen2000antenna}. A receiver that employs maximum ratio combining (MRC) chooses ${\bf z}$ such that ${\Gamma}({\bf f},{\bf z})$ for a given ${\bf f}$ is maximized~\cite{love2003grassmannian}. Under the assumption that receiver always uses MRC, ${\bf z}$ is given by 
\begin{align}
    {\bf z} &= {\bf H}{\bf f}/\norm{{\bf H}{\bf f}}_2,\notag
\end{align}
and ${\Gamma}({\bf f},{\bf z})$ can be simplified as
\begin{align}
    {\Gamma} := {\Gamma}({\bf f}) = {\norm{{\bf H}{\bf f}}^2_2}.\label{eq::beamforming::gain}
\end{align}
Therefore the MIMO beamforming problem is to find the optimal beamforming vector ${\bf f}$ that maximizes $\Gamma({\bf f})$ and can be formally posed as
\begin{align}
    {\bf f} &= \underset{{\bf x} \in \compvect} \argmax\ {\Gamma}({\bf x}).\label{eq::unconst::codeword::selection}
\end{align}
To constrain transmit power, we assume that $\norm{{\bf f}}^2_2 = 1$ without loss of generality. We consider maximum ratio transmission (MRT), which selects ${\bf f}$ to maximize ${\Gamma}({\bf f},{\bf z})$ for a given ${\bf z}$~\cite{lo1999maximum}. Under the assumptions of MRT, receive MRC, and no other design constraints on ${\bf f}$, for a given $N_o$ and ${\cal E}_t$, the optimal beamforming vector ${\bf f}$ that maximizes ${\Gamma}$ is 
\begin{align}
    {\bf f} &= \underset{{\bf x} \in \compvect}\argmax\ \norm{{\bf H}{\bf x}}^2_2 \text{ subjected to }  \norm{{\bf x}}_2^2 = 1 \notag \\
     &= \underset{{\bf x} \in \orthvect}\argmax\ \norm{{\bf H}{\bf x}}^2_2. \label{eq::const::codeword::selection}
\end{align}
Note that $\argmax$ of any function returns only one out of its possibly many global maximizers and thus the output may not necessarily be unique. For an MRT system, ${\bf f}$ is the orthonormal eigenvector associated with the maximum eigenvalue of ${\bf H}^H{\bf H}$~\cite{tse2000performance}. Let ${\lambda}_1 \geq \dots \geq {\lambda}_{M_t}$ be the eigenvalues of ${\bf H}^H{\bf H}$ and ${\bf v}_1,\dots,{\bf v}_{M_t}$ be the corresponding eigenvectors. One possible solution of \eqref{eq::const::codeword::selection} is ${\bf f} = {\bf v}_1$ and the corresponding beamforming gain is
\begin{align}
 {\Gamma}({\bf v}_1) &= \norm{{\bf H}{\bf v}_1}_2^2 =\lambda_{1}. \label{eq::MRT::beamformer}
\end{align}
For a given ${\bf H}$, let the solution space of \eqref{eq::const::codeword::selection} be denoted as ${\cal S}_{\bf H} \subset \orthvect$. Then ${\bf v}_1 \in {\cal S}_{\bf H}$ and for every ${\bf f} \in {\cal S}_{\bf H}$, $\Gamma({\bf f}) = \lambda_1$. 

\begin{figure}
    \centering
    \includegraphics[width = 0.45\textwidth]{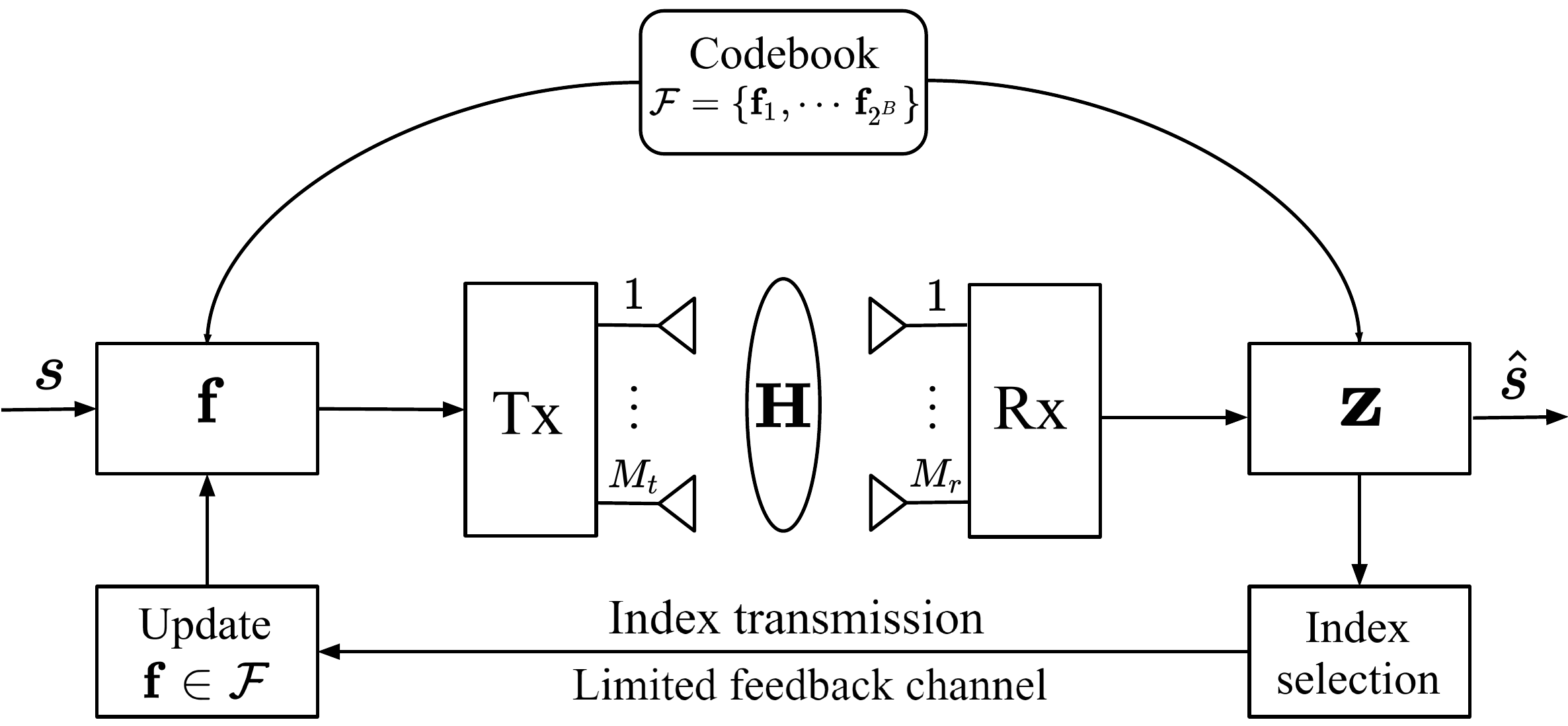}
    \caption{Block diagram of an FDD-MIMO system with limited feedback channel of capacity $B$ bits per channel use.}
    \label{fig::BlockDiagram}
\end{figure}

Quite obviously, MRT beamforming requires CSIT. In particular, in an FDD system, the receiver estimates the channel ${\bf H}$ and sends  ${\bf v}_1$ back to the transmitter over a feedback channel. Thus the feedback overhead increases as $M_t$ increases. Since the feedback channel is typically assumed to be a low-rate reliable channel, it is not always possible to transmit ${\bf v}_1$ over this channel without any data compression~\cite{narula1998efficient}. One way to model this feedback bottleneck is to assume the feedback channel to be a zero-delay, error-free, and the capacity being limited to $B$ bits per channel use. Thus, it is necessary to introduce some method of quantization for ${\bf v}_1$. The most well-known approach for the quantization is to construct a dictionary of beams \cite{narula1998efficient}, also known as the {\em beam codebook}. In particular, the transmitter and receiver agree upon a finite set of possible beamforming vectors, say ${\cal F} = \{{\bf f}_1, \dots ,{\bf f}_{2^B}\}$ of cardinality $2^B$. The receiver chooses the appropriate vector ${\bf f} \in {\cal F}$ that maximizes $\Gamma$ and feeds the index of the codeword back to the transmitter. The system-level diagram of a limited feedback FDD-MIMO system, as discussed so far, is provided in Fig.~\ref{fig::BlockDiagram}.  
Therefore for a given codebook $\cal F$, the optimal beamforming vector as stated in \eqref{eq::unconst::codeword::selection} is 
\begin{align}
 {\bf f} &=  \underset{{{\bf f}_i\in {\cal F}}} \argmax\ \norm{{\bf H}{\bf f}_i}^2_2. \label{eq::codeword::selection}
\end{align}
It is important to note that the original problem of finding the optimal MRT solution for \eqref{eq::const::codeword::selection} is a constrained optimization problem on the Euclidean space $\C^{M_t\times 1}$ and does not have unique solution on $\C^{M_t\times 1}$. This problem can be reformulated as a manifold optimization problem as follows. As argued in~\cite{love2003grassmannian}, it can be shown that the optimal MRT beamformers for every ${\bf H} \in {\C}^{M_r \times M_t}$ lie on a special kind of Riemann manifold embedded in $\C^{M_t\times 1}$, known as the Grassmann manifold. This will be discussed in Section~\ref{sec::Grassmannian}. As it will be evident later, this manifold structure of the search domain of ${\bf f}$ in \eqref{eq::const::codeword::selection} is the key enabler for our data-driven codebook design.

\section{Clustering on a Grassmann Manifold}\label{sec::Grassmannian}
In this section, we provide a brief introduction to the Grassmann manifold, which is instrumental for the design of the proposed beamforming codebook design. The reader is referred to the foundational texts in differential geometry, such as~\cite{boothby1986introduction}, for a comprehensive and rigorous treatment of this manifold. A Grassmann manifold refers to a set of subspaces embedded
in a higher-dimensional space (such as the surface of a sphere in $\mathbb{R}^3$). More formally, the complex Grassmann manifold ${\cal G}(M_t,M)$ is defined as 
\begin{align}
    {\cal G}(M_t,M) &:= \{\text{span}({\bf Y}) : {\bf Y} \in \C^{M_t \times M}, {\bf Y}^H{\bf Y} = {\bf I}_M\}.
\end{align}
Any element ${\cal Y}$ in ${\cal G}(M_t,M)$ is typically represented by an orthonormal matrix ${\bf Y}$ whose columns span ${\cal Y}$. It is to be noted that there exists no unique representation of a subspace ${\cal Y}$. This can be explained as follows. Let $\bf Y$ be the orthonormal basis that spans $\cal Y$, then ${\cal Y}$ can also be spanned by some other orthonormal matrix ${\bf Y}'={\bf Y}{\bf R}$ for some ${\bf R}\in {\cal U}_M$. Thus ${\bf Y}$ and ${\bf Y}'$ span the same subspace, which is represented by an equivalence relation ${\bf Y}' \equiv {\bf Y}$. Each of these $M$-dimensional linear subspaces can be regarded as a single point on the Grassmann manifold, which is represented by its orthonormal basis. Since a linear subspace can be specified by an arbitrary basis, each point on ${\cal G}(M_t,M)$ is an equivalence classes of orthonormal matrices. Specifically, ${\bf Y}$ and ${\bf Y}{\bf R}$ correspond to the same point on ${\cal G}(M_t,M)$. 

\begin{figure}[t]
    \centering
    \includegraphics[width = 0.25\textwidth]{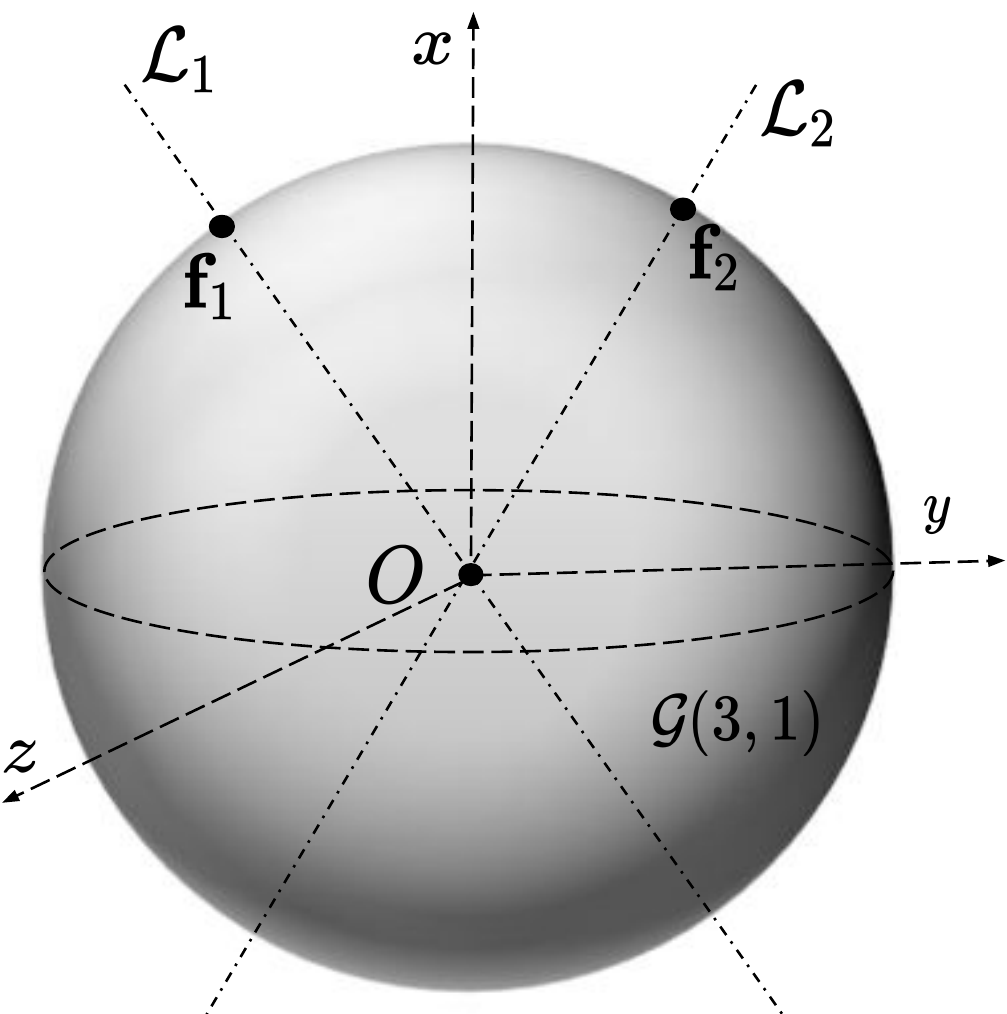}
    \caption{Example of a Grassmann manifold ${\cal G}(3,1)$ embedded in $\R^3$: ${\cal L}_1$, ${\cal L}_2$ are two arbitrary lines passing through the origin in $\R^3$. They are represented by unit vectors ${\bf f}_1$, ${\bf f}_2$, at the intersections of the lines ${\cal L}_1$, ${\cal L}_2$ and the surface of the unit sphere in $\R^3$.}
    \label{fig::Grassmannian::Manifold}
\end{figure}

Now consider the case of  $M = 1$, i.e. $\grvect$, which is the set of all one-dimensional subspaces in $\compvect$. In other words, one can visualize $\grvect$ as the collection of all lines passing through the origin of the space $\compvect$. A line ${\mathcal L}$ passing through origin in $\compvect$ is represented in $\grvect$ by a unit vector ${\bf f}$ that spans the line. It can also be generated by any other unit vector ${\bf f}^{\prime}$ if ${\bf f}^{\prime} \equiv {\bf f}$. Also note, ${\bf f}$ and ${\bf f}^{\prime}$ correspond to the same point on $\grvect$. Since a complex Grassmann of arbitrary dimensions is difficult to visualize, for illustration purpose, we present the real Grassmann manifold ${\cal G}(3,1)$ in $\R^3$ in Fig.~\ref{fig::Grassmannian::Manifold}. 

The notion of ``distance'' between the lines in ${\cal G}(M_t,1)$ generated by two unit vectors ${\bf f}_1, {\bf f}_2$ can be defined as the sine of the angle between the lines~\cite{conway1996packing}. In particular, the distance is expressed as
\begin{align}
     d({\bf f}_1, {\bf f}_2) &= \sin(\theta_{1,2}) = \sqrt{1-|{\bf f}^H_1{\bf f}_2|^2}. \label{eq::def::distance}
\end{align}
The connection between $\grvect$ and optimal MRT beamforming is established next.

\begin{lemma}\label{lem::Gr::uniqueness}
For a given channel realization ${\bf H} \in \C^{M_r \times M_t}$ and ${\bf f} \in {\cal S}_{\bf H}$, every ${\bf f}^{\prime}$ such that ${\bf f}^{\prime} \equiv {\bf f}$ is also an optimal MRT beamformer. 
\end{lemma}
\begin{IEEEproof}
Given ${\bf f} \in {\cal S}_{\bf H}$ and ${\bf f}^{\prime} \equiv {\bf f}$ i.e. ${\bf f}^{\prime} = {\bf f}e^{j\theta}$ for some $\theta \in [0, 2\pi)$, then 
\begin{align}
    \norm{{\bf H}{\bf f}}_2^2 &= |{\bf f}^H{\bf H}^H{\bf H}{\bf f}| \notag \\
    &= |e^{-j\theta}{\bf f}^H{\bf H}^H{\bf H}{\bf f}e^{j\theta}| \notag \\
    &= |{\bf f}'^H{\bf H}^H{\bf H}{\bf f}'| \notag \\
    &= \norm{{\bf H}{\bf f}'}_2^2.
\end{align}

Therefore, ${\bf f}^{\prime}$ is also an element in ${\cal S}_{\bf H}$. 
\end{IEEEproof}

\begin{remark}\label{rem::Gr::uniqueness}
According to Lemma~\ref{lem::Gr::uniqueness}, every ${\bf f}, {\bf f}^{\prime} \in {\cal S}_{\bf H}$ such that ${\bf f}^{\prime} \equiv {\bf f}$ correspond to the same point on $\grvect$. Therefore, any probability distribution on channel ${\bf H}\in \C^{M_r\times M_t}$ will impose a probability distribution on ${\bf v}_1$ in ${\cal G}(M_t,1)$. Thus we need a quantizer defined on ${\grvect}$ to encode the optimal MRT beamforming vectors and generate a $B$-bit beamforming codebook ${\cal F} \subset \grvect$ for limited feeback beamforming. 
\end{remark}
The optimal MRT beamformers are points on $\grvect$, hence can be described as a {\em point process} whose characteristics depend on the underlying channel distribution.  
\begin{remark}
For a Rayleigh fading channel where the entries of ${\bf H}$ are i.i.d according to ${\cal CN}(0,1)$, it has been shown in \cite{love2003grassmannian} that ${\bf v}_1$ is uniformly distributed on ${\cal G}(M_t,1)$. Thus, the construction of ${\cal F}$ is equivalent to finding the best packing of ${2^B}$ lines in ${\cal G}(M_t,1)$~\cite{conway1996packing}. See \cite{love2003grassmannian} for more details. 
\end{remark}

\begin{figure}
    \centering
    \includegraphics[width=0.75\linewidth]{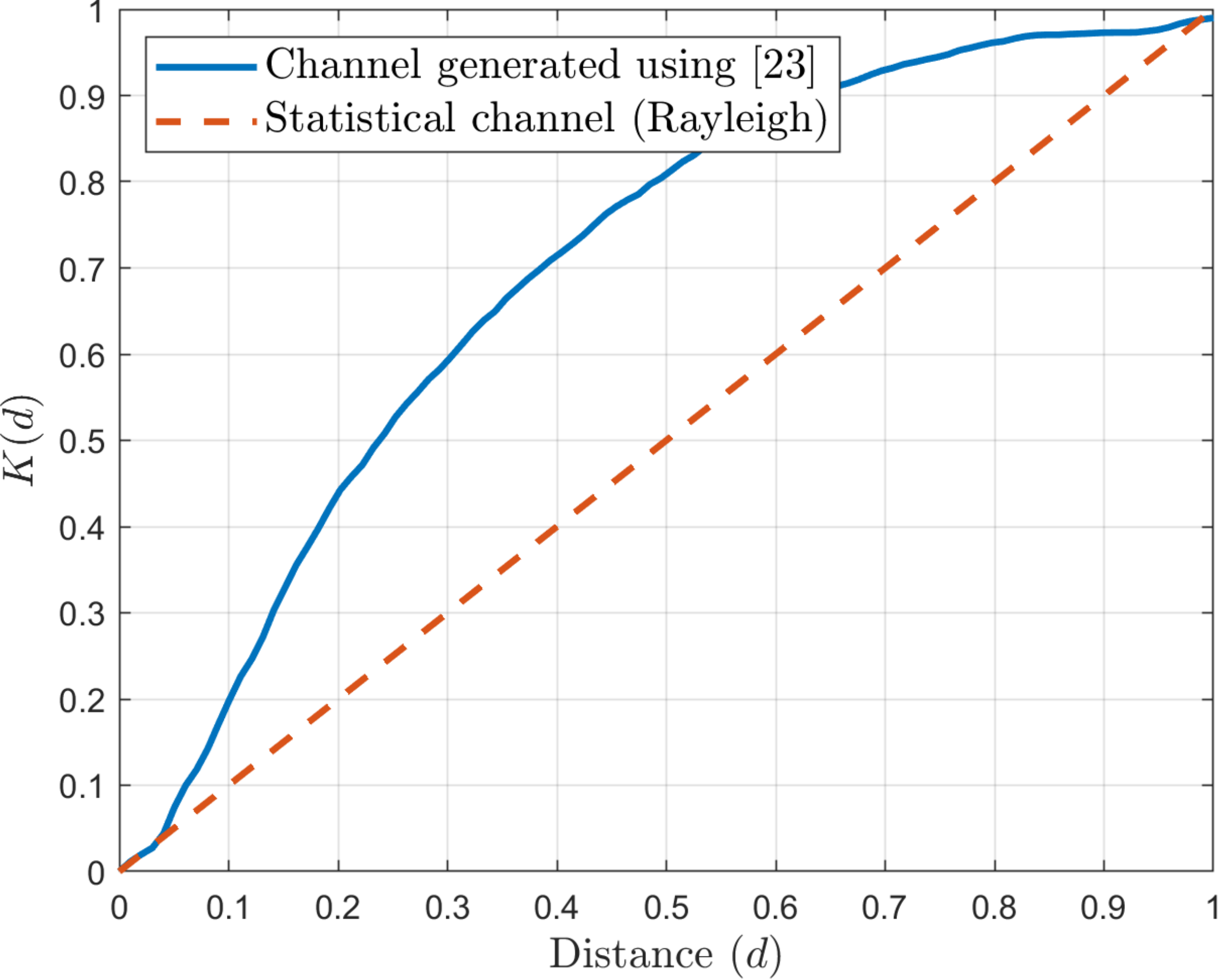}
    \caption{Comparison of Ripley's K function $K(d)$ of different channel models with $M_t = 2$ and $M_r = 1$.}
    \label{fig::RipleyK}
\end{figure}

For an arbitrary distribution of ${\bf H} \in \C^{M_r \times M_t}$, the distribution of ${\bf v}_1$ will no longer be uniformly distributed on ${\cal G}(M_t,1)$. As an illustrative example, using the distance function defined in $\eqref{eq::def::distance}$, we compare the Ripley's $K$ function $K(d)$~\cite{dixon2012ripley} of the optimal MRT beamforming vectors on $\grvect$ for a realistic scenario (see Section~\ref{sec::Results} for more details on the experimental setup) with Rayleigh fading channels for the same system model. Ripley's $K$ function is a spatial descriptive statistic that measures the deviation of a point process from spatial homogeneity. From Fig.~\ref{fig::RipleyK}, we see that the distribution of the optimal MRT beamforming vectors for the realistic channel is significantly different from the uniform distribution on $\grvect$ which is equivalent to {\em complete randomness} on the manifold. We can also infer that optimal MRT beamformers of the channel for the considered scenario exhibit clustering tendency on $\grvect$. Therefore, a reasonable codebook construction scheme is to simply deploy an unsupervised clustering method (such as $K$-means clustering) on ${\cal G}(M_t,1)$ that can identify optimal $K=2^{B}$ cluster centroids and form the codebook $\cal F$. {\em In what follows, we will establish that $K$-means clustering on ${\cal G}(M_t,1)$ actually yields an optimal codebook.}


\subsection{Grassmannian $K$-means clustering}
$K$-means clustering on a given metric space is a method of vector quantization to partition a set of $n$ data points into $K$ non-overlapping clusters in which each data point belongs to the cluster with the nearest cluster centroid. The centroids are the quantized representations of the data points that belong to the respective clusters. A quantizer on the given metric space maps the data points to one of the $K$ centroids. The $K$ centroids are chosen such that the average distortion due to the quantization according to a pre-defined distortion measure is minimized. Before we formally introduce the main steps of the clustering algorithm on $\grvect$, we first define the notion of a distortion measure and a quantizer as follows.

\begin{ndef}[Distortion measure]\label{def::distortion::function}
The distortion caused by representing ${\bf f} \in \grvect$ with ${\bf f}'\in \grvect$ is defined as the distortion measure $d_o$ which is given by $d_o({\bf f},{\bf f}^{\prime}) = d^2({\bf f},{\bf f}')$.
\end{ndef}
\begin{ndef}[Grassmann quantizer]\label{def::CSI::quantizer}
Let ${\cal F} \subset \grvect$ be a $B$-bit codebook such that ${\cal F} = \{{\bf f}_1,....,{\bf f}_{2^B}\}$, then a Grassmann quantizer $ Q_{\cal F}$ is defined as a function mapping elements of $\grvect$ to elements of ${\cal F}$ i.e. $Q_{\cal F}: \grvect \mapsto {\cal F}$.
\end{ndef}
A performance measure of a Grassmann quantizer is the average distortion $D(Q_{\cal F})$, where
\begin{align}
    D(Q_{\cal F}) &:= \E_{\bf x}\ \left[d_o\big({\bf x},Q_{\cal F}({\bf x})\big)\right] \notag \\ 
    &= \E_{\bf x}\ \left[d^2\big({\bf x},Q_{\cal F}({\bf x})\big)\right]. \label{eq::expectation::distortion}
\end{align} 
In most practical settings, we may have access to a set of $n$ data points ${\cal X} = \{{\bf x}\}$ in lieu of the probability distribution $p({\bf x})$. Then the expectation w.r.t ${\bf x}$ in \eqref{eq::expectation::distortion} means averaging over the set ${\cal X}$. Therefore the objective of $K$-means clustering with $K = 2^B$ is to find the set of $K$ centroids, i.e. ${\cal F}^{K}$, that minimizes $D(Q_{\cal F})$ and can be expressed as
\begin{align}
    {\cal F}^{K} &=  \underset{{\substack{{\cal F}\subset \grvect \\ |{\cal F}| = 2^B}}}\argmin\ D(Q_{\cal F}) \notag \\
    &= \underset{{\substack{{\cal F}\subset \grvect\\
     |{\cal F}| = 2^B}}}\argmin\ \E_{{\bf x}} \bigg[ d^2({\bf x},Q_{\cal F}({\bf x}))\bigg],\label{eq::Kmeans::objective}
\end{align}
and the associated quantizer is
\begin{align}
 Q_{{\cal F}^{K}}({\bf x}) &= \underset{{\bf f}_i \in {\cal F}}\argmin\ d_o({\bf x},{\bf f}_i) \notag \\
 &= \underset{{\bf f} \in {\cal F}}\argmin\ d^2({\bf x},{\bf f}_i).  \label{eq::CSI::quantizer}
\end{align}
However, finding the optimal solution for $K$-means clustering is an NP-hard problem. Therefore, we use the Linde-Buzo-Gray algorithm \cite{1094577} (outlined in Alg.~\ref{alg::Kmeans::Algo}) which is a heuristic algorithm that iterates between updating the cluster centroids and mapping a data point to the corresponding centroid that guarantees convergence to a local optimum. In Alg.~\ref{alg::Kmeans::Algo}, the only non-trivial step is the centroid calculation for a set of points. In contrast to the squared distortion measure in the Euclidean domain, the centroid of $n$ elements in a general manifold with respect to an arbitrary distortion measure does not necessarily exist in a closed form. However, the centroid computation on ${\cal G}(M_t,1)$ is feasible because of the following Lemma.
\begin{algorithm}
\caption{Grassmannian $K$-means Algorithm for \eqref{eq::Kmeans::objective}}\label{alg::Kmeans::Algo}
\begin{algorithmic}[1]
\Procedure{Codebook}{${{\cal X}},K$}
\State Initialize random ${\cal F} = \{{\bf f}_1,\cdots,{\bf f}_K\}$ on $\grvect$
\State  ${\cal V}_k \leftarrow \{{\bf x}: d({\bf x},{\bf f}_k) \leq d({\bf x},{\bf f}_j), \forall {\bf x} \in {\cal X}, k \neq j \}\ \forall k \in \{1,\cdots,K\}$
\State  $Q_{\cal F}({\bf x}$) $\leftarrow$ $\underset{{\bf f} \in {\cal F}}\argmin\ d^2({\bf x},{\bf f})\ \forall {\bf x} \in {\cal X}$
\While{$\text{!~stopping criteria}$}
\State Solve for ${\cal F}$:
\begin{subequations}\label{eq::Kmeans}
\begin{alignat}{2}
&{\bf f}_k \leftarrow \underset{\bf f}\argmin \sum d^2({\bf x},{\bf f})\ \forall {\bf x} \in {\cal V}_k, \forall k \in \{1,\cdots,K\}\notag\\
& Q_{\cal F}({\bf x}) \leftarrow \underset{{\bf f} \in {\cal F}}\argmin\ d^2({\bf x},{\bf f})\ \forall {\bf x} \in {{\cal X}}\notag\\
&{\cal V}_k \leftarrow \{{\bf x}: d({\bf x},{\bf f}_k) \leq d({\bf x},{\bf f}_j), \forall {\bf x} \in {\cal X}, k \neq j \}\ \forall k \in \notag \\
& \{1,\cdots,K\}\notag
\end{alignat}
\end{subequations}
\EndWhile
\Return ${\cal F}$
\EndProcedure
\end{algorithmic}
\end{algorithm} 
\begin{lemma}[Centroid computation]For a set of points ${\cal V}_k = \{{{\bf x}_i}\}^{N_k}_{i=1}$, ${\bf x}_i \in \grvect$, that form the $k$-th Voronoi partition, the centroid ${\bf f}_k$ is
\begin{align}
{\bf f}_k &= \underset{{\bf f} \in \grvect}\argmin \sum^{N_k}_{i=1}d^2({\bf x}_i,{\bf f}) = {\rm eig}\bigg(\sum^{N_k}_{i=1}{\bf x}_i{\bf x}^H_i\bigg), 
\end{align}
where ${\rm eig}({\bf Y})$ is the dominant eigenvector of the matrix ${\bf Y}$.
\end{lemma}

\section{Grassmannian Codebook Design}\label{sec::codebook::design}
In this section, we formally establish the connection between Grassmannian $K$-means clustering and the optimal codebook construction. The transmitter and receiver use a $B$-bit codebook ${\cal F} \subset \grvect$ to map the channel matrix ${\bf H}$ to a codeword in ${\bf f} \in {\cal F}$ according to \eqref{eq::codeword::selection}. In order to define the {\em optimality} of the codebook, we first introduce the average normalized beamforming gain for ${\cal F}$ as  
\begin{align}
{\Gamma}_{av} :&=  \E_{{\bf H}}\ \bigg[\frac{{\Gamma({\bf f})}}{\Gamma({\bf v}_1)}\bigg] \notag \\
&=  \E_{{\bf H}}\ \bigg[\frac{\norm{{\bf H}{\bf f}}_2^2}{\lambda_1}\bigg] \notag \\
&=  \E_{{\bf H}}\ \bigg[\sum^{M_t}_{i=1} \frac{{\lambda}_i|{\bf v}^H_i{\bf f}|^2}{{\lambda_1}}\bigg] \notag \\
&\geq  \E_{{\bf v}_1}\ \bigg[|{\bf v}^H_1{\bf f}|^2\bigg]. \notag
\end{align}
To measure the average distortion introduced by the quantization using ${\cal F}$, we use the loss in ${\Gamma}_{av}$ as given below:
\begin{align}
L({\cal F}) &:=  \E_{{\bf H}}\ \big[1 -{\Gamma}_{av}\big] \notag \\
&\leq  \E_{{\bf v}_1}\ \big[1 - |{\bf v}^H_1{\bf f}|^2\big] := L_{\rm ub}({\cal F}), \label{eq::BF::loss::UB}
\end{align}
where $L_{\rm ub}({\cal F})$ is an upper bound of $L({\cal F})$ and the sufficient condition for equality in~\eqref{eq::BF::loss::UB} is ${\rm rank}({\bf H}) = 1$. The optimal codebook intends to minimize $L({\cal F})$ by minimizing its upper bound $L_{\rm ub}({\cal F})$ as given in \eqref{eq::BF::loss::UB}. Since the current limited feedback approach quantizes ${\bf v}_1$ as ${\bf f}$, it is reasonable to minimize $L_{\rm ub}({\cal F})$ which depends only on ${\bf v}_1$ and ${\bf f}$. This yields the following codebook design criterion.
\begin{ndef}[Codebook design criterion]\label{def::codebook::design}
Over all of the $B$-bit codebooks ${\cal F} \subset \grvect$, the Grassmannian codebook ${\cal F}^*$ is the one that minimizes $L_{\rm ub}(\cal{F})$. Therefore
\begin{align}
  {\cal F}^* &:= \underset{{\substack{{\cal F}\subset \grvect \\ |{\cal F}| = 2^B}}}\argmin\ L_{\rm ub}({\cal F}). \label{eq::codebook::design::criteria}
\end{align}
\end{ndef}
Building on this discussion, we now state the method to find the optimal codebook in ${\cal G}(M_t,1)$ as follows. 
\begin{theorem}\label{thm::main}
For a feedback channel with capacity $B$ bits per channel use, the Grassmannian codebook as defined in Definition~\ref{def::codebook::design} is the same as the set of cluster centroids found by the $K$-means algorithm with $K = 2^B$ that minimizes $D(Q_{\cal F})$ for a given distribution of optimal MRT beamforming vector through its training dataset as given in \eqref{eq::expectation::distortion}, i.e. 
\begin{align}
    {\cal F}^* &= {\cal F}^{K}, \label{eq::codebook::design::Kmeans}
\end{align}
where ${\cal F}^{K}$ is given by \eqref{eq::Kmeans::objective}. 
\end{theorem}
\begin{IEEEproof}The optimal codebook is given by
\begin{align}
 {\cal F}^* &= \underset{{\substack{{\cal F}\subset \grvect \\ |{\cal F}| = 2^B}}}\argmin\ L_{\rm ub}({\cal F}) \notag \\
 &= \underset{{\substack{{\cal F}\subset \grvect \\ |{\cal F}| = 2^B}}}\argmin\ E_{{\bf v}_1}\ \big[1 - |{\bf v}^H_1{\bf f}|^2\big]\notag\\
&= \underset{{\substack{{\cal F}\subset \grvect \\ |{\cal F}| = 2^B}}}\argmin\ E_{{\bf v}_1}\ \bigg[\underset{{{\bf f}_i\in {\cal F}}} \min \big(1 - |{\bf v}^H_1{\bf f}_i|^2\big)\bigg] \notag\\ 
&= \underset{{\substack{{\cal F}\subset \grvect \\ |{\cal F}| = 2^B}}}\argmin\ E_{{\bf v}_1}\ \bigg[\underset{{\bf f}_i \in {\cal F}}\min\ d^2({{\bf v}_1,{\bf f}_i})\bigg] = {{\cal F}^K}, \notag
\end{align}
which completes the proof.
\end{IEEEproof}
 Theorem~\ref{thm::main} states the equivalence of the optimal codebook design with the $K$-means clustering on ${\cal G}(M_t,1)$. The benefit of making this connection is that it provides an approach for finding the optimal codebooks leveraging existing work on $K$-means clustering on $\grvect$. We are now in a position to state the key steps in designing the optimal codebooks based on the Grassmanian $K$-means clustering. 
 
 \subsection{Codebook Construction}
 We assume a stationary distribution of the channel for a given coverage area of a transmitter. In order to construct the Grassmannian codebook, we construct ${\cal H}=\{{\bf H}\}$, a set of channel realizations sampled for different user locations. The available channel dataset ${\cal H}$ is split into training and testing datasets, ${\cal H}_{\rm train}$ and ${\cal H}_{\rm test}$ for generating beamforming codebooks and evaluating their performance, respectively. We assume that the size of the training set is large enough so that the sampling distribution closely approximates the original distribution. The training procedure yields the optimal codebook whose performance is evaluated by measuring ${\Gamma}_{av}$ for the channel realizations in the test set ${\cal H}_{\rm test}$. Further details and benchmarking results are outlined in Section~\ref{sec::Results}. The codebook design and performance evaluation processes are illustrated in Alg.~\ref{alg:TrainingTesting::Algo}.
\begin{algorithm}
\caption{Training and testing of the Grassmannian codebook}\label{alg:TrainingTesting::Algo}
\begin{algorithmic}[1]
\Procedure{TrainCodebook}{$\cal H_{\rm train}$, $B$}
\State Initialize training set ${\cal X}_{\rm train}=\emptyset$ on ${\cal G}(M_t,1)$
\For{${\bf H}\in {\cal  H}_{\rm train}$}
\State ${\bf U}{\bf \Sigma}{\bf V}^H \leftarrow {\rm svd}({\bf H})$
\State ${\cal X}_{\rm train}\leftarrow {\cal X}_{\rm train}\cup {\bf v}_1 $
\EndFor 
\State ${\cal F}^*\leftarrow$ {\sc CodeBook}(${\cal X}_{\rm train}$,$K = 2^B$)

\Return ${\cal F}^*$
\EndProcedure
\Procedure{TestCodebook}{$\cal H_{\rm test}$, ${\cal F}^*$}
\State Initialize $\Gamma_{\rm av}=0$
\For{${\bf H}\in {\cal  H}_{\rm test}$}
\State ${\bf U}{\bf \Sigma}{\bf V}^H \leftarrow {\rm svd}({\bf H})$
\State ${\bf f}\leftarrow \underset{{{\bf f}'\in {\cal F}^*}} \argmin\ d^2({\bf v}_1,{\bf f}')$
\State $\Gamma_{\rm av} \leftarrow  \Gamma_{\rm av}+\frac{1}{\#{\cal H}_{\rm test}} \bigg(\frac{\Gamma({\bf f})}{\Gamma({\bf v}_1)}\bigg)$
\EndFor 
\Return ${\Gamma}_{\rm av}$
\EndProcedure
\end{algorithmic}
\end{algorithm}

\section{Grassmannian Product Codebook for FD-MIMO}\label{sec::product::codebook::design}
After discussing the general notion of the codebook construction for transmit beamforming for an $M_t\times M_r$ MIMO system, we focus our attention to a special case of FD MIMO communication. We assume that the transmitter is equipped with a UPA with dimensions $M_v\times M_h$ ($M_hM_v = M_t$) while the receiver has one antenna, i.e. $M_r = 1$. Let $\tilde{\bf H}$ represent the $M_v \times M_h$ channel matrix where the $(i,j)$-th element corresponds to the channel between the antenna element at the $i$-th row and $j$-th column of the UPA and the single receiver antenna at the user. Note that this system model appears as a special case of the general $M_t\times M_r$ MIMO system discussed in the previous section where ${\bf H}^T={\rm vec}(\tilde{\bf H}^T)$. Hence the codebook can be designed as given in  Alg.~\ref{alg:TrainingTesting::Algo} using the $K$-means clustering in ${\cal G}(M_v M_h,1)$. Assuming $M_v=M_h = O(n)$, the dimension of the codewords increase as $O(n^2)$. Naturally, the $K$-means clustering will suffer from the {\em curse of dimensionality} as the difference in the maximum and minimum distances between two points in the dataset becomes less prominent as the dimension of the space increases~\cite{friedman2001elements}. In this section, we show that the codebook can be obtained by clustering on lower dimensional manifolds by exploring the geometry of the UPA. Considering the UPA channel as a matrix $\tilde{\bf H}$, we have the singular value decomposition of $\tilde{\bf H}$ as follows. 
\begin{align}
    \tilde{\bf H} &= {\bf U}{\bf \Sigma}{\bf V}^H, \label{eq::UPA::channel}
\end{align}
where ${\bf U}$ is the left singular matrix, ${\bf V}$ is the right singular matrix and ${\bf \Sigma}$ is the rectangular diagonal matrix with singular values $\sigma_i$ in decreasing order. Let ${\lambda}_i$ be the $i$-th eigenvalue of $\tilde{\bf H}^H\tilde{\bf H}$, then ${\lambda}_i = {\sigma}^2_i$. Further, ${\bf u}_i$ and ${\bf v}_i$ are the column vectors of ${\bf U}$ and ${\bf V}$ respectively with  ${\bf u}^H_i{\bf u}_i = 1$ and ${\bf v}^H_i{\bf v}_i = 1$. Thus,  ${\bf u}_i \in {\cal U}(M_v,1)$ and ${\bf v}_i \in {\cal U}(M_h,1)$. Then we have
\begin{align}
    {\bf H}^T &= {\rm vec}({\tilde{\bf H}}^T) \notag \\
    &= {\rm vec}({\bf V}^*{\bf \Sigma}{\bf U}^T)\notag\\
                &= {\rm vec}\bigg(\sum^{{\rm rank}(\tilde{\bf H})}_{i=1} \sigma_i{\bf v}_i^*{\bf u}^T_i\bigg) \notag \\
                &= \sum^{{\rm rank}(\tilde{\bf H})}_{i=1} \sigma_i{\bf u}_i\otimes {\bf v}^*_i. \label{eq::h::decomp}
\end{align}
From \eqref{eq::h::decomp}, we can represent ${\bf H}$ as the linear combination of ${\bf u}^T_i\otimes {\bf v}^H_i$ scaled with $\sigma_i$. Thus we have
\begin{align}
      {\bf H} &= \sum^{{\rm rank}(\tilde{\bf H})}_{i=1} \sigma_i{\bf u}^T_i\otimes {\bf v}^H_i. \label{eq::h::KP::decomp}
\end{align}
 Due to the finiteness of the physical paths between the transmitter to the receiver, it is well-known that ${\rm rank}({\bf H})<<\min(M_v,M_h)$. For the sake of simplicity, we approximate the channel ${\bf H}$ with its dominant direction, i.e. ${\bf u}^T_1\otimes {\bf v}^H_1$, which is called ${\rm rank}$-$1$ approximation and the approximated channel $\bar{\bf H}$ is given as 
\begin{align}
{\bf H} &\approx \bar{\bf H} = {\sigma_1}{\bf u}^T_1\otimes {\bf v}^H_1. \label{eq::rank1::approx}
\end{align}
Let ${\bf x} \in \orthvect$ be a beamforming vector for $\bar{\bf H}$. Then the KP form of $\bar{\bf H}$ naturally leads us to the idea of using ${\bf x}$ of the form ${\bf x} = {\bf x}_v\otimes {\bf x}_h$. This motivates us to use separate codebooks ${\cal F}_h$ and ${\cal F}_v$ for the horizontal and vertical dimensions and enables to design product codebooks by clustering in lower dimensional manifolds. The beamforming gain ${\Gamma}({\bf x})$ for $\bar{\bf H}$ can now be simplified as 
\begin{align}
{\Gamma}({\bf x}) &= \norm{\bar{\bf H}{\bf x}}^2_2\notag \\
&=  \norm{{\sigma_1}({\bf u}^T_1\otimes {\bf v}^H_1)({\bf x}_v\otimes {\bf x}_h)}^2_2 \notag \\
&\stackrel{(a)}{=} {\sigma^2_1}\norm{{\bf u}^T_1{\bf x}_v}_2^2\ \norm{{\bf v}^H_1{\bf x}_h}_2^2 \notag \\
&=  {\sigma^2_1}\ |{\bf u}^T_1{\bf x}_v|^2\ |{\bf v}^H_1{\bf x}_h|^2, \notag
\end{align}
where step $(a)$ follows from the fact that $\norm{{\bf A}\otimes {\bf B}}_2 = \norm{{\bf A}}_2\norm{{\bf B}}_2$ for two matrices ${\bf A}$ and ${\bf B}$ of any dimensions. The optimal MRT beamforming vector ${\bf f}$ for $\bar{\bf H}$ can be simplified as 
\begin{align}
{\bf f} &= \underset{{\bf x} \in {\orthvect}}\argmax\ { \Gamma({\bf x})} \notag \\
&= \underset{{\substack{{\bf x}_v \in {\cal U}(M_v,1) \\ {\bf x}_h \in {\cal U}(M_h,1)}}} \argmax\ |{\bf u}^T_1{\bf x}_v|^2\ |{\bf v}^H_1{\bf x}_h|^2 \notag \\
&=  \underset{{\bf x}_v \in {\cal U}(M_v,1)} \argmax\ |{\bf u}^T_1{\bf x}_v|^2 \notag \\
&\qquad \otimes \underset{{\bf x}_h \in {\cal U}(M_h,1)} \argmax\ |{\bf v}^H_1{\bf x}_h|^2 \notag \\
&= {\bf f}_v \otimes {\bf f}_h,  \label{eq::prod::codeword}
\end{align}
where
\begin{align}
{\bf f}_v = \underset{{\bf x}_v \in {\cal U}(M_v,1)} \argmax\ |{\bf u}^T_1{\bf x}_v|^2, \ {\bf f}_h =  \underset{{\bf x}_h \in {\cal U}(M_h,1)} \argmax\ |{\bf v}^H_1{\bf x}_h|^2. 
\end{align}
Observe that one possible solution for optimal MRT beamformer ${\bf f}$ in \eqref{eq::prod::codeword} is given by ${\bf f}_v = {\bf u}^*_1$ and ${\bf f}_h = {\bf v}_1$. Following Remark~\ref{rem::Gr::uniqueness}, we can argue that ${\bf f}={\bf f}_v \otimes {\bf f}_h$, where ${\bf f}_{u}\in{\cal G}(M_v,1)$ and ${\bf f}_v\in{\cal G}(M_h,1)$. 

The loss in average normalized beamforming gain ${\Gamma_{av}}$ with the codebook ${\cal F} = {\cal F}_v \otimes {\cal F}_h$ can be bounded as 
\begin{align}
 L({\cal F}) &= \E_{{\bf u}_1, {\bf v}_1}\ \bigg[1- \frac{{\Gamma({\bf f}_v \otimes {\bf f}_h)}}{\Gamma({\bf u}^*_1 \otimes {\bf v}_1)}\bigg] \notag \\
&= \E_{{\bf u}_1, {\bf v}_1}\ \bigg[1 - |({\bf u}^T_1\otimes {\bf v}^H_1)({\bf f}_v\otimes {\bf f}_h)|^2 \bigg] \notag\\
&\leq 2 \E_{{\bf u}_1, {\bf v}_1}\ \bigg[1 - |({\bf u}^T_1\otimes {\bf v}^H_1)({\bf f}_v\otimes {\bf f}_h)| \bigg] \notag\\
&\leq  2 \E_{{\bf u}_1, {\bf v}_1}\ \bigg[\underset{\theta, \phi}\min \big( \norm{(e^{j\theta}{\bf u}^*_1 \otimes e^{j\phi}{\bf v}_1) - ({\bf f}_v \otimes {\bf f}_h)} \big) \bigg] \notag\\
&\leq 2 \E_{{\bf u}_1, {\bf v}_1}\ \bigg[\underset{\theta, \phi}\min \big(\norm{e^{j\theta}{\bf u}^*_1}_2\norm{e^{j\phi}{\bf v}_1 - {\bf f}_h}_2 + \notag \\
&\qquad \qquad \norm{e^{j\theta}{\bf u}^*_1 - {\bf f}_v}_2\norm{e^{j\phi}{\bf f}_h}_2 \big) \bigg] \notag\\
&= 2 \E_{{\bf u}_1, {\bf v}_1}\ \bigg[ \underset{\theta, \phi}\min \big(\norm{e^{j\phi}{\bf v}_1 - {\bf f}_h}_2 + \norm{e^{j\theta}{\bf u}^*_1 - {\bf f}_v}_2 \big) \bigg] \notag\\
&= 2 \E_{{\bf u}_1, {\bf v}_1}\ \bigg[(1 - |{\bf v}^H_1{\bf f}_h|)^{1/2} + (1 - |{\bf u}^T_1{\bf f}_v|)^{1/2} \bigg] \notag\\
&\leq\scalebox{0.9}{$ 2 \E_{{\bf u}_1, {\bf v}_1}\ \bigg[(1 - |{\bf v}^H_1{\bf f}_h|^2) + (1 - |{\bf u}^T_1{\bf f}_v|^2) \bigg]$}= L_{\rm ub}({\cal F}). \label{eq::UPA::BF::loss::UB}
\end{align}

\begin{ndef}[Grassmannian product codebook]\label{def::grassmann::product::codebook}
Under the rank-1 approximation of the channel, ${\bf H} \approx \bar{\bf H} = \sigma_1{\bf u}^T_1 \otimes {\bf v}^H_1$, the $B$-bit Grassmannian product codebook ${\cal F}^* = {\cal F}_v^* \otimes {\cal F}_h^*$ is the one that satisfies the codebook design criteria in Definition~\ref{def::codebook::design} for a given $[B_v, B_h]$ where $|{\cal F}^*_v| = 2^{B_v}$, $|{\cal F}^*_h| = 2^{B_h}$ and $B_h + B_v = B$.
\end{ndef}
We will now state the method to construct the product codebook ${\cal F}^*$ as follows.
\begin{lemma}\label{lem::prod::codebook::design}
The Grassmannian product codebook ${\cal F}^* = {\cal F}_v^* \otimes {\cal F}_h^*$ as defined in Definition~\ref{def::grassmann::product::codebook} is constructed using the set of centroids ${\cal F}^K_v$ and ${\cal F}^K_h$ obtained from the independent $K$-means clustering of the optimal MRT beamforming vectors ${\bf v}_1$ and ${\bf u}^*_1$ on ${\cal G}(M_v,1)$ and ${\cal G}(M_h,1)$ with $K = 2^{B_v}$ and $K = 2^{B_h}$, respectively. 
\end{lemma}
\begin{IEEEproof}
From Definition~\ref{def::grassmann::product::codebook}, 
\begin{align*}
{\cal F}^* &= {\cal F}^*_v \otimes {\cal F}^*_h \notag \\
&= \underset{{\cal F}_v, {\cal F}_h}\argmin\ L_{\rm ub}({\cal F})\\
&=\underset{{\cal F}_v, {\cal F}_h}\argmin\ \E_{{\bf u}_1, {\bf v}_1}\ \bigg[(1 - |{\bf v}^H_1{\bf f}_h|^2) + (1 - |{\bf u}^*_1{\bf f}_v|^2) \bigg]\\
&= \underset{{\substack{{\cal F}_v\subset {\cal G}(M_v,1) \\ |{\cal F}_v| = 2^{B_v}}}}\argmin \E_{{\bf u}_1}\ (1 - |{\bf u}^T_1{\bf f}_v|^2)\ \otimes \notag \\
&\qquad \qquad \underset{{\substack{{\cal F}_h\subset {\cal G}(M_h,1) \\ |{\cal F}_h| = 2^{B_h}}}}\argmin \E_{{\bf v}_1}\ (1 - |{\bf v}^H_1{\bf f}_h|^2),
\end{align*}
\begin{align}
\text{where }
{\cal F}^*_v &= \underset{{\substack{{\cal F}_v \subset {\cal G}(M_v,1)\\ |{\cal F}| = 2^{B_v}}}}\argmin\ \E_{{\bf u}_1}\ \big[1 - |{\bf u}^T_1{\bf f}_v|^2 \big],\notag \\
{\cal F}^*_h &= \underset{{\substack{{\cal F}_h\subset {\cal G}(M_h,1) \\ |{\cal F}| = 2^{B_h}}}}\argmin\E_{{\bf v}_1}\ \big[1 - |{\bf v}^H_1{\bf f}_h|^2 \big].\notag
\end{align}
Following Theorem~\ref{thm::main}, we have ${\cal F}_h^* = {\cal F}_h^K, {\cal F}_v^* = {\cal F}_h^K$. Thus, the Grassmannian product codebook is given as
\begin{align}
    {\cal F}^* &= {\cal F}^*_v \otimes {\cal F}^*_h = {\cal F}^K_v \otimes {\cal F}^K_h.
\end{align}
\end{IEEEproof}

From Lemma~\ref{lem::prod::codebook::design}, it is possible to perform $K$-means clustering independently on ${\cal G}(M_v,1)$, ${\cal G}(M_h,1)$ and construct the product codebook with reduced complexity instead of performing $K$-means clustering on ${\cal G}(M_hM_v,1)$. The training and testing procedure of the proposed product codebook design for a given set of channel realizations ${\cal H}$ is given in the following remark.
\begin{remark}
For a training set ${\cal H}_{\rm train}$, the Grassmann product codebook ${\cal F}^*$ defined as ${\cal F}^* = {\cal F}^*_{v}\otimes {\cal F}^*_h$, where ${\cal F}^*_{v}$ and ${\cal F}^*_{h}$ are obtained by the procedure {\sc TrainProdCodeBook}(${\cal H}_{\rm train}$,$[B_v, B_h]$) using Alg.~\ref{alg:UPA::TrainingTesting::Algo}, where $B_v$, $B_h$ are the number of bits used to encode ${\bf u}^*_1$, ${\bf v}_1$ respectively ($B=B_v+B_h$). The performance of the codebook ${\cal F}^*$ is evaluated by the procedure {\sc TestProdCodeBook}(${\cal H}_{\rm test}$,$[{\cal F}_v^*, {\cal F}_h^*]$) as given in Alg.~\ref{alg:UPA::TrainingTesting::Algo}.
\end{remark}
\begin{algorithm}
\caption{Training and testing of the Grassmannian product codebook}\label{alg:UPA::TrainingTesting::Algo}
\begin{algorithmic}[1]
\Procedure{TrainProdCodebook}{$\cal H_{\rm train}$, $[B_v,B_h]$}
\State Initialize training sets ${\cal X}_{\rm train}=\emptyset$ and ${\cal Y}_{\rm train}=\emptyset$ on 

${\cal G}(M_h,1)$ and ${\cal G}(M_v,1)$ respectively
\For{${\bf H}\in {\cal  H}_{\rm train}$}
\State Generate $\tilde{\bf H}$ from ${\bf H}$, ${\bf U}{\bf \Sigma}{\bf V}^H \leftarrow {\rm svd}(\tilde{\bf H})$
\State ${\cal X}_{\rm train}\leftarrow {\cal X}_{\rm train}\cup {\bf v}_1 $, ${\cal Y}_{\rm train}\leftarrow {\cal Y}_{\rm train}\cup {\bf u}^*_1 $
\EndFor 
\State ${\cal F}_h^*\leftarrow$ {\sc CodeBook}(${\cal X}_{\rm train}$, $2^{B_h}$)
\State ${\cal F}_v^*\leftarrow$ {\sc CodeBook}(${\cal Y}_{\rm train}$, $2^{B_v}$)

\Return $[{\cal F}_v^*$, ${\cal F}_h^*]$
\EndProcedure
\Procedure{TestProdCodebook}{$\cal H_{\rm test}$, $[{\cal F}_v^*$, ${\cal F}_h^*]$}
\State Initialize $\Gamma_{\rm av}=0$
\For{${\bf H}\in {\cal  H}_{\rm test}$}
\State Generate $\tilde{\bf H}$ from ${\bf H}$, ${\bf U}{\bf \Sigma}{\bf V}^H \leftarrow {\rm svd}(\tilde{\bf H})$
\State ${\bf f}_h \leftarrow \underset{{{\bf f}'\in {\cal F}_h^*}} \argmin\ d^2({\bf v}_1, {\bf f}')$, 
\State ${\bf f}_v \leftarrow \underset{{{\bf f}'\in {\cal F}_v^*}} \argmin\ d^2({\bf u}^*_1, {\bf f}')$
\State $\Gamma_{\rm av} \leftarrow  \Gamma_{\rm av}+\frac{1}{\#{\cal H}_{\rm test}} \frac{\Gamma({\bf f}_v \otimes {\bf f}_h)}{\Gamma({\bf u}^*_1 \otimes {\bf v}_1)}$
\EndFor
\Return ${\Gamma}_{\rm av}$
\EndProcedure
\end{algorithmic}
\end{algorithm} 
\section{Simulations and Discussions}\label{sec::Results}
The proposed learning framework requires channel datasets to construct the corresponding beamforming codebooks. In this section, we describe the scenario adopted for the channel dataset generation and evaluate the performance of the generated codebooks in terms of ${\Gamma}_{av}$.
\subsection{Dataset generation}\label{sec::dataset::gen}
In this paper, we consider an indoor communication scenario between the base station and the users with $M_r = 1$ operating at a frequency of 2.5 GHz. The described scenario is a part of the DeepMIMO dataset \cite{alkhateeb2019deepmimo}. The channel dataset with the parameters given in Table.~\ref{dataset::table} is generated using the dataset generation script of DeepMIMO. 
\begin{table}
\caption{Parameters of the DeepMIMO dataset}\label{dataset::table}
\centering
\begin{tabular}{ |c|c| } 
 \hline
 Name of scenario & I1\_2p5 \\ 
  \hline
 Active BS & 3  \\ 
  \hline
 Active users & 1 to 704  \\ 
 \hline
Number of antennas (x, y, z) & ($M_v,M_h,1$) \\
 \hline
 System bandwidth & 0.2 GHz \\
 \hline
 Antennas spacing & 0.5 \\
 \hline
 Number of OFDM sub-carriers & 1 \\
 \hline
 OFDM sampling factor & 1 \\
 \hline
 OFDM limit & 1 \\
 \hline
\end{tabular}
\vspace{0.1in}
\end{table}
          
\subsection{Results}
We now demonstrate the performance of the proposed codebook design techniques through simulations. The dataset generated as described in Section~\ref{sec::dataset::gen} is represented as ${\cal H} = \{{\bf H}\}$ where ${\bf H} \in \mathbb{C}^{1 \times M_t}$ and partitioned into training and testing datasets, ${\cal H}_{\rm train}$ and ${\cal H}_{\rm test}$ respectively. We refer to the Grassmannian codebook design in Section~\ref{sec::codebook::design} as Prop. I : $[B]$ when $B$ bits are used for generating the codebook as illustrated in Alg.~\ref{alg:TrainingTesting::Algo}. The Grassmannian product codebook design described in Section~\ref{sec::product::codebook::design} is referred as Prop. II : $[B_v, B_h]$ where $[B_v, B_h]$ is the allocation of the feedback bits for constructing the codebooks using Alg.~\ref{alg:UPA::TrainingTesting::Algo} and $B = B_v + B_h$. 
 
We compare the average normalized beamforming gain ${\Gamma}_{av}$ of the two proposed codebooks with that of the DFT structured KP codebooks~\cite{choi2015advanced} (referred as KP DFT) and the codebooks generated based on the Grassmannian line packings for correlated channel~\cite{love2006limited} (referred to as GLP). The Grassmannian line packings required for the codebook construction in~\cite{love2006limited} were obtained from \cite{love2003grassmannian}, \cite{medra2014flexible}. The channel correlation matrix ${\bf R}$ is calculated from the training channel dataset ${\cal H}_{\rm train}$ according to ${\bf R} = \E_{\bf H}\big({{\bf H}^H{\bf H}}\big)$.

 \begin{figure}
  \centering
              \includegraphics[width=0.75\linewidth]{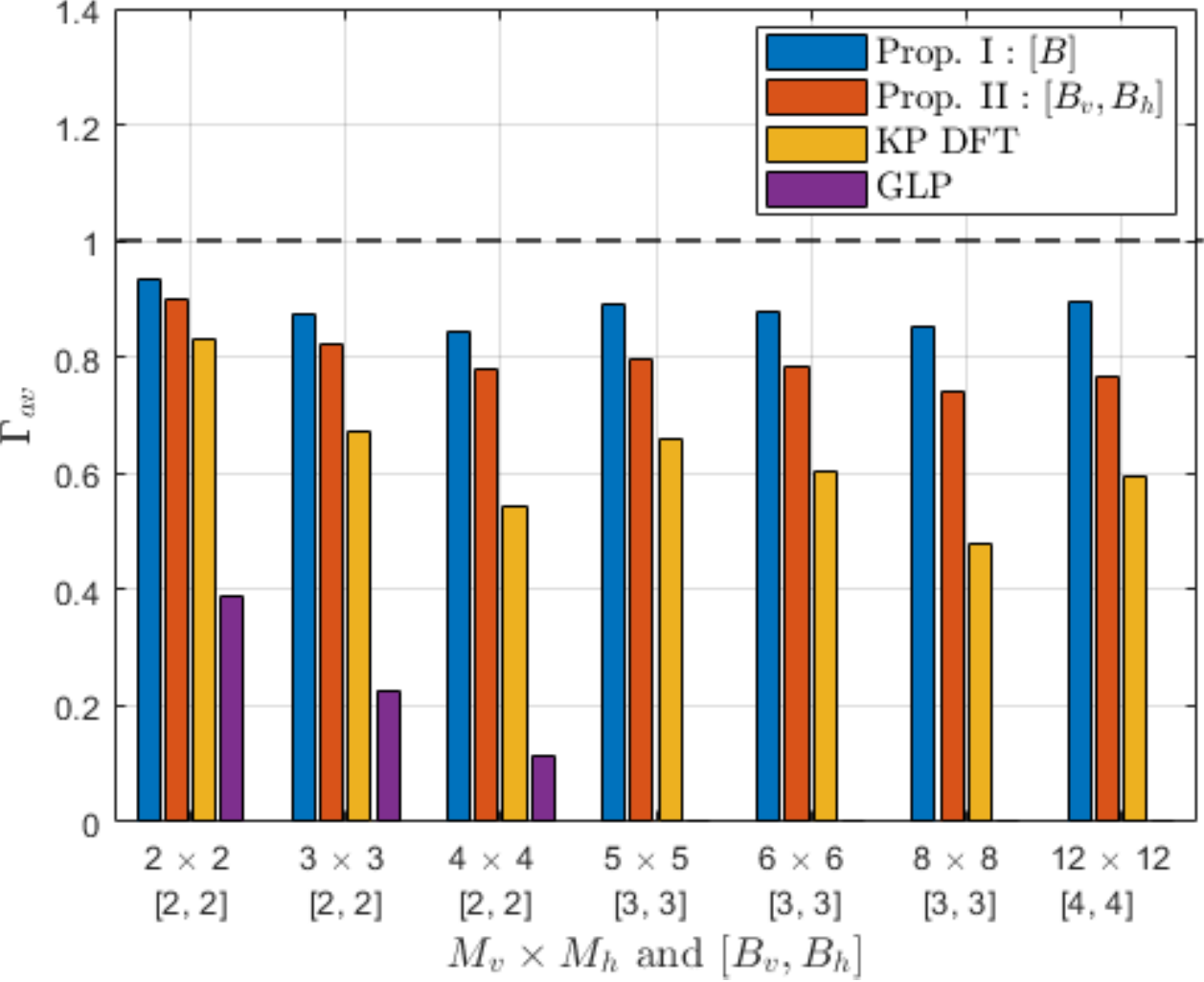} \\ \vspace{.1in}
              \includegraphics[width=0.75\linewidth]{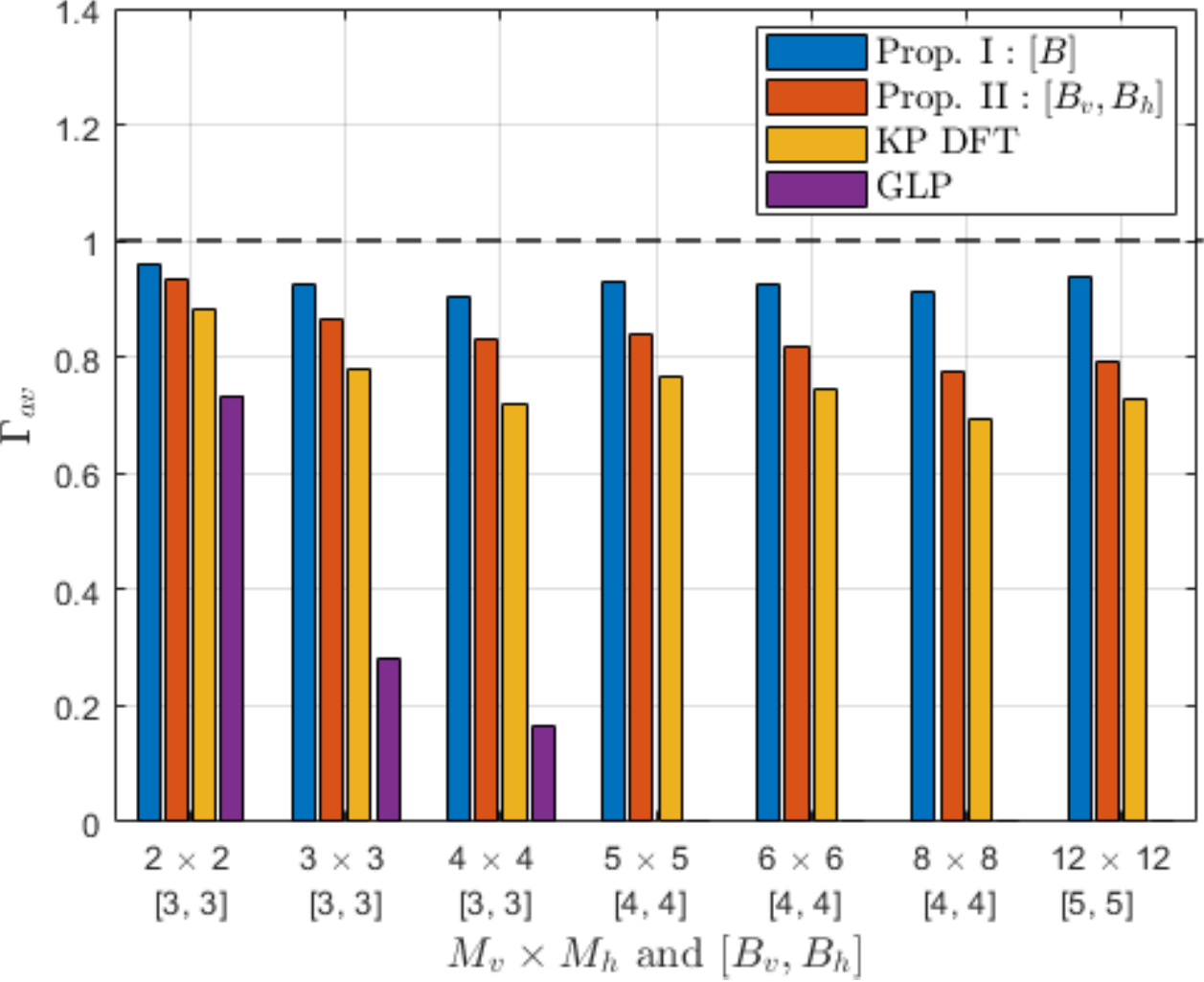}
 \caption{Average normalized beamforming gain for different transmit antenna configurations $M_v \times M_h$ and feedback-bit allocations $[B_v, B_h]$.}
 \label{fig::BFgain} 
 \end{figure} 
In Fig.~\ref{fig::BFgain}, the normalized beamforming gains for different transmit antenna configurations $M_v \times M_h$ and feedback-bit allocations $[B_v, B_h]$ using the four codebook design techniques are plotted. The KP codebooks relying on DFT structure are simple to construct but the quantized beams may not be effective because the KP codebooks search only the beams lying in the direction of the right and left dominant singular vectors of the reshaped channel in \eqref{eq::UPA::channel}. While the Grassmannian line packings were shown to be the optimal codebooks for uncorrelated Rayleigh fading channel and were extended to correlated Rayleigh fading channel, they do not necessarily work well with any arbitrary channel distribution. It was not possible to show the performance of the GLP codebooks for all $M_v \times M_h$ because of the challenges in finding the best line packings. Simulation results in Fig. \ref{fig::BFgain} demonstrate that the codebooks designed using the proposed techniques adapt well to the underlying channel distribution. They also outperform KP DFT codebooks~\cite{choi2015advanced}, GLP codebooks~\cite{love2006limited} and provide beamforming gains comparable to that of optimal MRT beamforming. Equivalently, they reduce the feedback overhead because they can maintain the same quantization performance with less overhead compared to the other codebooks.

\section{Conclusion}
This paper considered the problem of codebook based MRT beamforming in an FDD-MIMO system operating under arbitrary channel conditions. Leveraging well-known connections of this problem with Grassmannian line packing, we identified that the problem of finding the optimal MRT beamforming codebooks for any arbitrary channel distribution can be constructed using the $K$-means clustering of MRT beamforming vectors on $\grvect$. We presented the Grassmannian $K$-means clustering algorithm to construct the beamforming codebooks. We showed that this approach can be extended to FD-MIMO systems with UPA antennas to design product codebooks with reduced computational complexity. As our future work, we will be investigating the design of optimal codebooks for limited feedback unitary precoding for spatial multiplexing in MIMO systems with arbitrary underlying channel. Overall, this paper provides a concrete example of a design problem with rigorous mathematical underpinnings that benefits from a classical {\em shallow} learning approach.

\bibliographystyle{IEEEtran}  
\bibliography{Draft_arxiv.bbl}

\end{document}